\def\HESS{\mbox{H.E.S.S. }}

\newcommand{\rb}[1]{\raisebox{1.5ex}[-1.5ex]{#1}}

\documentclass[twocolumn,runningheads]{svjour2}
\smartqed  % flush right qed marks, e.g. at end of proof
\usepackage{graphicx}
\usepackage{mathptmx}
\journalname{Astrophysics and Space Science}
\begin{document}

\title{Observations of Extragalactic Sources with the MAGIC Telescope
}
\subtitle{TeV Blazars and Extragalactic Background Light}

\author{Daniel Mazin for the MAGIC collaboration
}

\institute{Daniel Mazin \at
              Max-Planck-Institut f\"ur Physik \\
              F\"ohringer Ring 6 \\
              D-80805 M\"unchen \\
              Tel.: +49-89-32354259\\
              Fax: +49-89-32354516\\
              \email{mazin@mppmu.mpg.de}           %  \\
}

\date{Received: date / Accepted: date}

\maketitle

\begin{abstract}
MAGIC is currently the world's largest single dish ground based imaging atmospheric
Cherenkov telescope. During the first year of operation, more than 20
extragalactic sources have been observed and several of them detected. Here we
present results of analyzed data, including discussion about spectral and
temporal properties of the detected sources. In addition, we discuss
implications of the measured energy spectra of distant sources for our knowledge
of the extragalactic background light.  
\end{abstract}

\section{Introduction}
\label{intro}
The search for very high energy (VHE, defined as $E\geq 100$ GeV) $\gamma$-ray
emission from Active Galactic Nuclei (AGNs) is one of the major goals for
ground-based $\gamma$-ray astronomy. New detections open up a possibility of
phenomenological studies of the physics inside the relativistic jets in AGNs, in
particular of understanding both the origin of the VHE $\gamma$-rays as well as
the relations between photons of different energy (from radio to VHE).  The
number of extragalactic $\gamma$-ray sources detected by EGRET on board the
CGRO with high confidence amounts to 66 \cite{hartman}.  However, the number 
of AGNs reported to be VHE $\gamma$-ray emitters (although slowly increasing)
is currently just thirteen (July 2006).

Stecker et al.\ \cite{ste} pointed out that the
attenuation of $\gamma$-rays due to photon-photon interactions with low
energy photons from the extragalactic background light (EBL) was a likely
explanation of this deficit.  In fact, the redshifts of blazars
detected so far above 100\,GeV have rather low values 
as expected from predictions of the correlation between the
gamma-ray attenuation and the redshift of the source, known as
Fazio-Stecker relation \cite{fazio,kneiske}. 

All known VHE gamma-ray emitting blazars belong to the class of
high-frequency-peaked BL Lacertae objects \linebreak (HBLs, \cite{padovani}), a
subclass of blazars characterized by a low luminosity when compared with
quasars and a synchrotron peak in the X-ray band. Their Spectral Energy
Distribution (SED) is characterized by a second peak at very high gamma-ray
energies. In synchrotron-self-Compton (SSC) models it is assumed that the
observed gamma-ray peak is due to the inverse-Compton (IC) emission from the
accelerated electrons up-scattering previously produced synchrotron photons to
high energies \cite{mar}.
A compilation of blazars with known X-ray spectra allowing their classification
as HBLs is given in \cite{don}.

The Major Atmospheric Gamma-ray Imaging Cheren-kov (MAGIC) telescope 
observed a sample of X-ray bright ($F_{1\,\rm{keV}}$\,$>$\,$2\,\rm{\mu Jy}$)
northern HBLs at moderate  redshifts ($z$\,$<$\,$0.3$).
The sample of candidates was chosen based on predictions from 
models involving an SSC \cite{CostGhis} and hadronic \cite{man2} 
origin of the $\gamma$-rays.

The known VHE $\gamma$-ray emitting AGNs are variable in flux in all
wavebands.  Correlations between X-ray and $\gamma$-ray \linebreak 
emission have been found on time scales ranging from several 10 minutes to days
(e.g.  \cite{fossati04}) although the relationship has proven to be rather
complicated.
The optical-TeV correlation has yet to be studied, but the optical-GeV
correlations seen in 3C~279 \cite{hartman01} suggest that at least in some
sources such correlations do exist. Using this as a guideline, the MAGIC
collaboration has performed Target of Opportunity observations whenever
they were alerted about sources being in a high flux state in the optical and/or
X-ray band. 

Here we present results of the detected extragalactic \linebreak sources ordered by
increasing redshift values: Mkn~421 \linebreak (z=0.030), Mkn~501 (z=0.034), 1ES\,2344+514
(z=0.044), Mkn~180 (z=0.045), 1ES\,1959+650 (z=0.047), 1ES\,1218+304 (z=0.182),
PG 1553+113 (unknown redshift, z$>$0.09 \cite{z_new}). The results from galactic
observations are presented elsewhere in these proceedings \cite{rico}.

%%%%%%%%%%%%%%%%%%%%%%%%%%%%%%%%%%%%%%%%%%%%%%%
%   MAGIC 
%%%%%%%%%%%%%%%%%%%%%%%%%%%%%%%%%%%%%%%%%%%%%%%

\section{MAGIC}
\label{sec:0}

\begin{figure*}
\centering
  \includegraphics[width=0.9\textwidth]{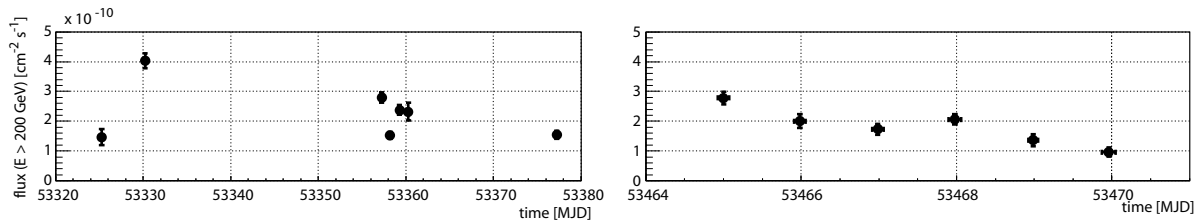}
\caption{ Lightcurve for Mkn~421 from November 2004
to April 2005. Left panel: data from November 2004 to January 2005.  Right
panel: data for April 2005. } 
\label{fig:421lc} 
\end{figure*}

The MAGIC telescope \cite{cortina} is located on the Canary Island La Palma
(2200 m asl, 28$^\circ$45$'$N, 17$^\circ$54$'$W) and is currently the
largest imaging atmospheric Cherenkov telescope worldwide. 
The accessible energy range
spans from 50-60 GeV (trigger threshold at small zenith angles) up to tens of
TeV. The sensitivity of MAGIC is $\sim$~2.5\% of the Crab Nebula flux in 50 hours
of observations. The energy resolution is about 30\% above 100 GeV and about 
25\% above 200 GeV.
The $\gamma$ point spread function is about $0.1$~degrees. 

The standard operation mode for MAGIC is the ON-observation, with the source
position in the center of the camera.  Usually, to get a robust estimate of the
background, part of the data are taken in the OFF mode, where a sky region is
tracked by the telescope which resembles the region around the source with
respect to the level of the night sky background fluctuations.  Part of the
data were taken in the so-called WOBBLE mode \cite{daum}.  In this mode, two
sky directions, opposite and 0.4$^\circ$ off source respectively, were tracked
alternately for 20 minutes each.  The advantage of the WOBBLE tracking mode is
a simultaneous measurement of the background, and thus {\em a~priori} no need
for additional OFF data. 

For calibration, image cleaning, cut optimization, and energy reconstruction,
the MAGIC standard analysis chain \cite{rwagner} was used.  For $\gamma$/hadron
separation as well as for energy determination a multidimensional
classification technique based on the Random Forest method \cite{breiman,rf}
was used.  The cuts were then chosen such that the overall cut efficiency for
Monte-Carlo $\gamma$ events was about 50$\%$. The systematic errors were
estimated to be around 50\% for the absolute flux level and 0.2 for the
spectral index.

In parallel with the observations of the AGNs with \linebreak MAGIC, the sources were observed
with the KVA 35~cm telescope (http://tur3.tur.iac.es/), also located on La Palma.  In
addition to these joint campaigns, several AGNs are regularly observed as
part of the Tuorla Observatory Blazar monitoring program with the Tuorla 1~m
and the KVA 35~cm telescopes.   

In the following sections we present the highlights of the extragalactic
observations by the MAGIC telescope. The observational details of the selected
sources including an analysis energy threshold are summarized in
Table~\ref{tab:1}.

\begin{table}
\caption{Summary of observation details of the presented data. 
ZA is the zenith angle in degrees.
$\mbox{E}_{\tiny \mbox{Thr}}$ is the analysis threshold in GeV.
T(hr) is the observation time in hours.}
\centering
\label{tab:1}       % Give a unique label
\begin{tabular}{lllll}
\hline\noalign{\smallskip}
{\bf Source} & {\bf Period} & {\bf ZA} & 
$\mbox{\bf E}_{\tiny \mbox{\bf Thr}}$ & {\bf T(hr)/Mode } \\[3pt]
\tableheadseprule\noalign{\smallskip}
               &  Nov04--Apr05  & 9--55 &           & 15.5/ON     \\
\rb {Mkn 421}  & Apr05          & 9--32 & \rb{150}  & 10.1/WOBBLE \\
\hline
{Mkn 501}   & {Jun--Jul05}  & 10--31 & {150} & 29.7/ON \\
\hline
1ES 2344 & Sep--Dec05 & 23--34 & 160 & 27.6/WOBBLE \\
\hline
Mkn 180  & Mar06 & 39--44 & 200 & 11.1/WOBBLE \\
\hline
1ES 1959 & Sep--Oct04 & 36--46& 180 & 6.0/ON \\
\hline
1ES 1218 & Jan05  & 2--13& 140 & 8.2/ON \\
\hline
         & Apr--May05   & 12-30 &           & 7.0/ON     \\
\rb{PG 1553} & Jan--Apr06 & 20-30 & \rb{140} & 11.8/ON \\
\noalign{\smallskip}\hline
\end{tabular}
\end{table}

%%%%%%%%%%%%%%%%%%%%%%%%%%%%%%%%%%%%%%%%%%%%%%%
%   Mrk 421 
%%%%%%%%%%%%%%%%%%%%%%%%%%%%%%%%%%%%%%%%%%%%%%%

\section{Markarian 421}
\label{sec:1}

Mkn~421 (redshift $z$ = 0.030) is the closest known TeV blazar and, 
along with Mkn~501, also the best studied one.  It was the first extragalactic $\gamma$-ray
source detected in the TeV energy range using IACTs \cite{punch}.
So far, Mkn~421 has shown variations larger than one order of
magnitude and occasional flux doubling times as short as 15
min~\cite{gaidos}.  Variations in the hardness of the TeV
$\gamma$-ray spectrum during flares were reported by several groups (e.g.
\cite{krenn,hess421}).  Simultaneous observations in the X-ray and GeV-TeV
bands showed a significant flux correlation \cite{krawczynski421}.  

Mkn~421 was observed with the MAGIC for a total of 19 nights, the observation
times per night ranging from 30 minutes up to 4 hours (Table~\ref{tab:1}).  Most
of the data were taken at small zenith angles ($ZA < 30^\circ$).  Only 1.5 hours in
December 2005 were taken at $42^\circ < ZA < 55^\circ$ during simultaneous
observations with  \HESS \cite{hessmagic}.  

\begin{figure}
\centering
  \includegraphics[width=0.31\textwidth,angle=0,clip]{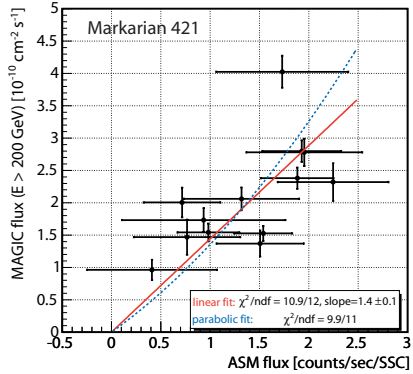}
\caption{
	Correlation plot between VHE $\gamma$-ray flux above 200 GeV 
    and X-ray counts for 11 nights of Mkn 421 data.}
\label{fig:421Xray}       % Give a unique label
\end{figure}

During the entire observation period Mkn~421 was found to be in a 
medium flux state ranging from 0.5 to 2 Crab units above 200~GeV (see
Fig.~\ref{fig:421lc}). Significant variations of up to a factor of four overall
and up to a factor two in between successive nights can be seen.
Fig.~\ref{fig:421Xray} shows a clear correlation between X-ray 
(taken from the All-Sky-Monitor on-board the RXTE satellite) and VHE
$\gamma$-ray data. The points are the nightly average of the MAGIC data and 
the simultaneous ASM count rate.
A linear fit as well as a parabolic fit are forced to go through
(0,0) and describe the correlation well. The linear correlation coefficient
is $ r = 0.64^{+0.15}_{-0.22}$.

\begin{figure}
\centering
  \includegraphics[width=0.42\textwidth,angle=0,clip]{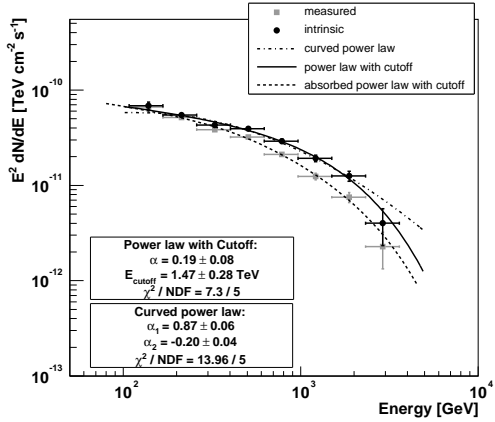}
\caption{
         The measured (grey points) and de-absorbed (black points) spectrum 
         of Mkn~421, multiplied by $E^2$. 
         Solid line: fit (1) to the intrinsic spectrum using a power 
         law with a cut-off. 
         Dashed-dotted line: fit (2) to the intrinsic spectrum using 
         a curved power law. A dashed line indicates the expected absorbed 
         spectrum using the result of fit (1). 
         Fit parameters of the intrinsic spectrum 
         are shown in the inlays.}
\label{fig:421spec}       % Give a unique label
\end{figure}

The measured spectrum and the reconstructed de- \linebreak absorbed (i.e. corrected for
the effect of extragalactic absorption) spectrum are shown in
Fig.~\ref{fig:421spec}. Here we used the recent model of 
Primack et al.\,\cite{primack}, which is in agreement with the upper limits set 
by H.E.S.S.\,\cite{hessebl} and low limits from the galaxy counts \cite{spitzer,elbaz}.
The de-absorbed
spectrum is clearly curved. Therefore, it is obvious that the curvature
in the measured spectrum has an intrinsic origin rather than being caused by the
absorption of the VHE $\gamma$-rays by the EBL photons.
A fit with a pure power law with a exponential cut-off as well as a fit 
with a curved power law indicate a flattening of the spectrum
towards 100~GeV.  
For details of the analysis and results see \cite{magic421}.

%%%%%%%%%%%%%%%%%%%%%%%%%%%%%%%%%%%%%%%%%%%%%%%
%   Mrk 501 
%%%%%%%%%%%%%%%%%%%%%%%%%%%%%%%%%%%%%%%%%%%%%%%

\section{Markarian 501}
\label{sec:2}

\begin{figure*}
\centering
  \includegraphics[width=0.80\textwidth,angle=0,clip]{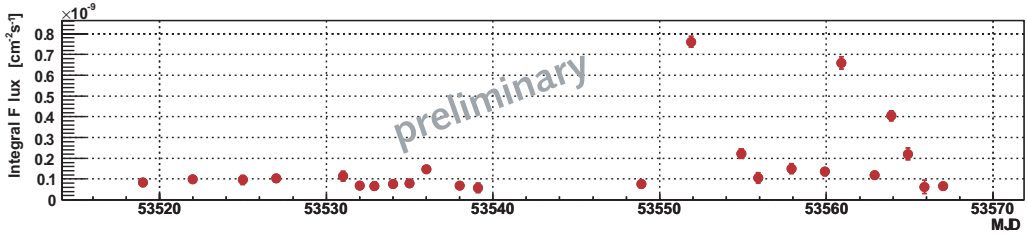}
\caption{ Lightcurve for Mkn~501 from June--July 2005. Shown are night averaged
 integral fluxes above 200 GeV.}
\label{fig:501lc}
\end{figure*}

The AGN Mkn 501 \cite{Quinn1996} is one of the best--studied objects in VHE
$\gamma$-rays. The source is known to be a strong and variable VHE $\gamma$-ray
emitter. During a flare
in 1997, Mkn~501 showed strong variability on
timescales of 0.5 days.  The integral flux reached 10
times the flux of the Crab nebula above $1$~TeV \cite{hegra501}. 
The  position of the IC peak was not detected yet while it was
observed that the spectrum gets harder with an increasing flux level. 

\begin{figure}
\centering
  \includegraphics[width=0.40\textwidth,angle=0,clip]{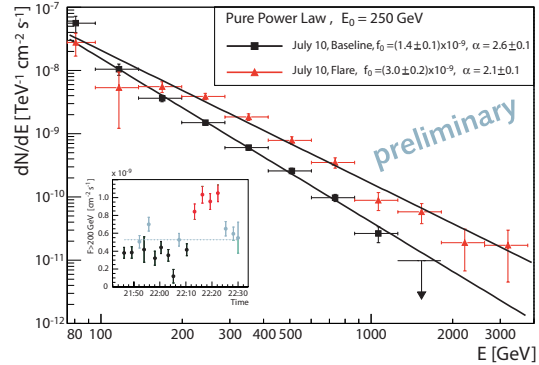}
\caption{Spectral differential energy distribution of Mkn~501 on 
10 July 2005. The rapid variation of the flux level and corresponding
change in the shape of the energy spectrum on a time scale of some 10 minutes
is clearly visible.}
\label{fig:501spe} 
\end{figure}

Mkn~501 was observed for 24 nights (Table~\ref{tab:1}) by the MAGIC telescope. 
It is worth noticing that $\sim$18 out of 30 hrs of the observation time
were performed in the presence of (moderate) moonshine. 
The source was found to be in a rather low flux state during most of the
observations. The flux level above 200~GeV was 30\%-50\% of the Crab Nebula
flux with a strong indication of an IC peak.  
On five nights, the source was found in a flaring state with
the flux reaching up to 4 Crab units (Fig~\ref{fig:501lc}).  Moreover, a rapid
flare with a doubling time as short as 5 minutes or less was detected on the
night of 10 July 2005 (inlay in Fig.~\ref{fig:501spe}). 
No change in background rates was seen during the observed flare.
The rapid increase in the flux level was accompanied by a hardening of the 
differential spectrum  (Fig.~\ref{fig:501spe}). This is the first time that
spectral hardening was detected on time scales of some 10 minutes.  A detailed
publication on the analysis and results on the observation of Mkn~501 is in
preparation.

%%%%%%%%%%%%%%%%%%%%%%%%%%%%%%%%%%%%%%%%%%%%%%%
%   1ES\,2344+514
%%%%%%%%%%%%%%%%%%%%%%%%%%%%%%%%%%%%%%%%%%%%%%%

\section{1ES\,2344+514}
\label{sec:3}

1ES\,2344+514 was identified as a BL~Lac object
by \cite{Perlman}, who also determined a redshift of $z=0.044$ from
absorption line measurements, while no evident emission lines
were found.  In VHE $\gamma$-rays, the source was observed by the
Whipple collaboration during the 1995/96 observing season,
yielding a signal on the $5.8\,\sigma$ level
\cite{Catanese1998}. Over 1/3 of the measured excess was recorded 
during the night of December 20, 1995, with corresponding photon flux 
above 350 GeV of 64\% of the Crab nebula flux.
The VHE $\gamma$-ray emission of 1ES\,2344+514 was confirmed on
a significance level of $4.4\,\sigma$ by the HEGRA
collaboration \cite{54AGN}, who found a quiescent flux
level approximately 50 times lower than during the 1995 flare.

The MAGIC observation of 1ES\,2344+514 (Table~\ref{tab:1}) \linebreak
yielded a clear excess with the significance of 11.5\,$\sigma$.
The measured flux corresponds to 6\% of the Crab Nebula flux.
No strong evidence for flux variability on times scales of 
days or weeks was found. The differential energy spectrum  
can be fitted by a simple power law with a photon index of $2.96 \pm 0.12$.

This was the first time that VHE $\gamma$-rays from 1ES 2344+ \linebreak 514 were 
detected with high significance in a
quiescent flux state. The derived spectrum is softer than the one 
reported by the Whipple collaboration during the flare in 1995 \cite{schroedter}.
A detailed publication on the analysis and results of MAGIC observations 
of 1ES\,2344+514 is in preparation.

%%%%%%%%%%%%%%%%%%%%%%%%%%%%%%%%%%%%%%%%%%%%%%%
%   Markarian 180
%%%%%%%%%%%%%%%%%%%%%%%%%%%%%%%%%%%%%%%%%%%%%%%

\section{Markarian 180}
\label{sec:4}

The AGN Mkn~180 (1ES 1133+704) is a well-known HBL at a redshift of $z=0.045$
\cite{falco}.  Previous attempts to detect VHE $\gamma$-rays only resulted in upper
limits \cite{54AGN,horan}.

\begin{figure}
\begin{center}
\includegraphics[width=0.40\textwidth,angle=0,clip]{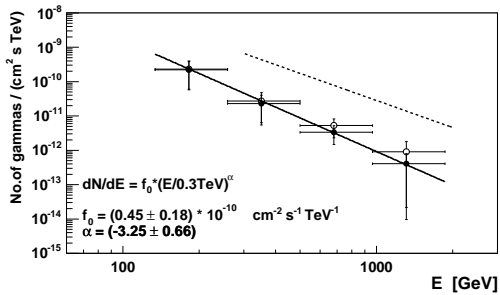}
\caption{\label{fig_spectrum} The differential energy spectrum of Mkn~180.
    Full circles: the spectrum measured by MAGIC. 
    Open circles: the de-absorbed energy spectrum (see text).
    The horizontal bars indicate the size of each energy bin.
    The black line represents a power law fit to the
    measured spectrum. The fit parameters are listed in the figure.
    For comparison, the Crab Nebula energy spectrum as derived from MAGIC 
    data \cite{rwagner} is shown (dashed line).}
\end{center}
\end{figure}

The observation of Mkn~180 was triggered by a brightening of the source in the
optical on March 23, 2006, detected by the KVA telescope.  The alert was given as
the core flux increased by 50\% from its quiescent level value.
Mkn~180 was observed by the MAGIC telescope in 2006 during 8 nights (Table~\ref{tab:1}).
The signal of 165 excess events 
was found with a significance of 5.5\,$\sigma$.
No evidence for flux variability between nights was found.  
The measured energy spectrum of Mkn~180 is shown in Fig.~\ref{fig_spectrum}.
A fit by a power law gives a photon index 
$\alpha=3.3\pm0.7$.
The observed integral flux above 200~GeV is
11\% of the Crab Nebula flux.  
The attenuation of the spectrum caused by the EBL was determined by
numerical integration of Eq.~2 in \cite{dwek}.  The de-absorbed energy
spectrum of Mkn~180 is also shown in Fig.~\ref{fig_spectrum} (open circles).  A
fit with a simple power law to the de-absorbed spectrum gives a slope with
$\alpha^{\prime}=2.8\pm0.7$. 

The discovery of VHE emission from Mkn~180 during an optical outburst makes it
very tempting to speculate about the connection between optical activity and
increased VHE emission.  Since Mkn~180 has not been observed with \linebreak MAGIC prior
to the outburst and the upper limits from other experiments are above the
observed flux level, further observations are needed.  For details of this
analysis and results see \cite{magic180}.

%%%%%%%%%%%%%%%%%%%%%%%%%%%%%%%%%%%%%%%%%%%%%%%
%   1ES\,1959+650
%%%%%%%%%%%%%%%%%%%%%%%%%%%%%%%%%%%%%%%%%%%%%%%

\section{1ES\,1959+650}
\label{sec:5}

The Seven Telescope Array in Utah reported for the first time
a VHE $\gamma$-ray signal from 1ES 1959+650 in 1998  
with the significance of 3.9\,$\sigma$ \cite{seven}. 
In May 2002, when the X-ray flux of the source had significantly
increased, both the Whipple~\cite{holder} and HEGRA~\cite{hegra1} collaborations  
subsequently confirmed 1ES 1959+650 as a VHE $\gamma$-ray source.
An interesting aspect of the source activity in 2002 was the observation 
of a so-called {\it orphan flare}, i.e. a VHE $\gamma$-ray activity is observed 
in the absence of high activity in X-rays.
Orphan flare in VHE $\gamma$-rays are not expected within the SSC mechanism 
in relativistic jets. 

1ES 1959+650 was observed by MAGIC during the co-missioning
phase in 2004 (Table~\ref{tab:1}).
The analysis of the data gave a detection of VHE $\gamma$-rays 
on a significance of 8.2\,$\sigma$. 
We obtained an integral VHE $\gamma$ flux above 180~GeV of
\linebreak
(3.73~$\pm$0.41)~$\cdot$10$^{-11}$ ph~cm$^{-2}$~s$^{-1}$. 
The energy spectrum between 180 GeV and 2~TeV 
is compatible with a power law with a photon index 
$\alpha=\mbox{2.72}\pm\mbox{0.14}$ and
is consistent with the quiescent state measured by HEGRA \cite{hegra1}.
This is the first time 1ES 1959+650 has been observed down to 180~GeV. 
For the details of the analysis and results see \cite{magic1959}.

%%%%%%%%%%%%%%%%%%%%%%%%%%%%%%%%%%%%%%%%%%%%%%%
%   1ES\,1218+304
%%%%%%%%%%%%%%%%%%%%%%%%%%%%%%%%%%%%%%%%%%%%%%%

\section{1ES\,1218+304}
\label{sec:6}

The AGN 1ES\,1218+304 was observed several times 
with the Whipple telescope between 1995 and 2000. The observations
resulted in an upper flux limit above 350\,GeV  of \linebreak
$8.3 \cdot 10^{-12}$\,ph~cm$^{-2}$~s$^{-1}$ 
corresponding to
$\sim$8\% of the Crab Nebula flux \cite{horan}. The source was
also observed 
by HEGRA between 1996 and 2002 and an upper flux limit above 840 GeV of $2.67
\cdot 10^{-12}$\,ph~cm$^{-2}$~s$^{-1}$     
(or 12\% of the Crab Nebula flux) was reported \cite{54AGN}.

\begin{figure}
\begin{center}
\includegraphics[width=0.40\textwidth,angle=0,clip]{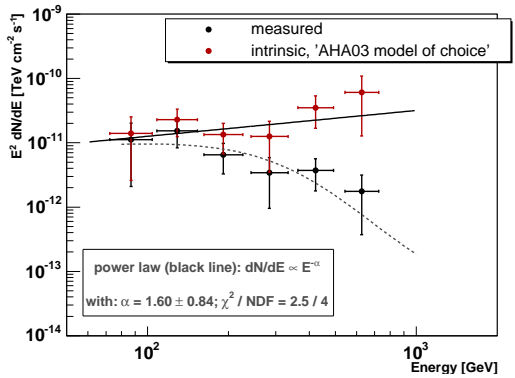}
\caption{Impact of the EBL on the differential energy spectrum of 
1ES\,1218+304.
{\it Black points}: measured spectrum.
{\it Red points}: reconstructed intrinsic spectrum of 1ES\,1218+304
using the EBL density "model of choice" from \cite{aha01}, 
which is clearly excluded in \cite{hessebl}. 
{\it Black line}: pure power law fit to the intrinsic spectrum,
the fit parameters are listed in the inlay.} 
\label{fig:1218spec}
\end{center}
\end{figure}

MAGIC observed 1ES\,1218+304 in seven nights in January 2005 
(Table~\ref{tab:1}).
The observed excess of 560 events has a
statistical significance of 6.4 standard deviations above 140 GeV.
1ES\,1218+304 is the first source discovered by MAGIC.
The nightly averaged $\gamma$-ray lightcurve did not show signs of 
significant variability.
The energy spectrum was fitted with a pure power law with a
photon index $\alpha=3.0\pm0.4$, and 
the determined flux level is below the upper limits at higher
energies determined in the past.
For details of the analysis and results see \cite{magic1218}.

Measured VHE $\gamma$-ray spectra of distant sources can be used to derive
constraints on the EBL density.  Recently, using the VHE $\gamma$-ray spectra
of two new detected HBL's at a similar redshift as 1ES 1218+304, H.E.S.S. derived
an upper limit on the EBL density between 1 and 4 $\mu$m \cite{hessebl}.
The main assumption needed to derive the limit is that the photon index of the intrinsic
(de-absorbed) spectrum of an AGN can not be harder than 1.5.  Though under
debate \cite{katar}, this assumption remains conservative if the VHE
$\gamma$-rays are produced in shock regions.  We did the same exercise using
the 1ES 1218+304 spectrum. However, our result does not constrain the EBL
density that strong compared to the upper limit that has been recently derived
by H.E.S.S. due to large statistical errors of the measured spectral points and
the fact that MAGIC did not measure flux points above 700 GeV where the effect 
of the EBL becomes very significant.
Fig.~\ref{fig:1218spec} illustrates the effect assumimg a relative high EBL density
on the 1ES 1218+304 spectrum. Though the de-absorbed spectrum 
is very hard, the fitted photon index $\alpha = 1.6 \pm 0.8$ is within the allowed
range ($>$1.5).

%%%%%%%%%%%%%%%%%%%%%%%%%%%%%%%%%%%%%%%%%%%%%%%
%   PG\,1553+113
%%%%%%%%%%%%%%%%%%%%%%%%%%%%%%%%%%%%%%%%%%%%%%%

\section{PG\,1553+113}
\label{sec:7}

\begin{figure}
\begin{center}
\includegraphics[width=0.40\textwidth,angle=0,clip]{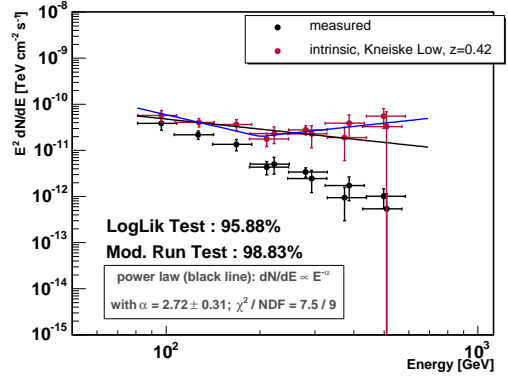}
\caption{Constraint on the redshift of PG 1553+113. 
{\it Black points}: measured combined differential energy spectrum of PG 1553+113 using 
MAGIC and H.E.S.S. data from 2005 and 2006. 
{\it Red points}: reconstructed intrinsic spectrum of PG 1553+113, using
minimum possible density of the evolving EBL and the redshift of z=0.42.
{\it Black line}: power law fit to the intrinsic spectrum; 
the fit parameters are listed in the inlay.
{\it Blue line}: broken power law fit.}
\label{fig:1553spec}
\end{center}
\end{figure}

PG 1553+113 belongs to a catalog of X-ray bright objects \cite{don} and,
based on its SED properties, was one of the most promising candidates from a
list of VHE $\gamma$-ray emitting \linebreak AGNs proposed by \cite{CostGhis}.
So far, upper limits on the $\gamma$-ray emission have been reported by the
Whipple collaboration (19\% Crab flux above 390 GeV) \cite{deperez03} and
Milagro \cite{williams04}. Recently the H.E.S.S. collaboration has presented
evidence for a $\gamma$-ray signal at the $4\,\sigma$ level (up to $5.3\,\sigma$
using a low energy threshold analysis) above 200 GeV corresponding to about 2\%
of the Crab flux \cite{hess1553}. 

PG 1553+113 was observed with the MAGIC telescope in 2005 at 
about the same time as the H.E.S.S. observations took place.
Motivated by a hint of a signal in the preliminary analysis of the MAGIC data, 
additional observations were performed in 2006 (Table~\ref{tab:1}).
Combining the data from 2005 and 2006, a very clear signal  
was detected with a total significance of $8.8\,\sigma$. 
In $\gamma$-rays there was no evidence for
short term variability on a time scale of days, but a significant change by a factor of three 
in the flux level from 2005 to 2006 was found. 
The combined 2005 and 2006 differential energy spectrum for PG
1553+113 is well described by a pure power law with a photon index
$\alpha=4.2\pm0.4$. 

The signal detected by MAGIC confirms the tentative signal seen by H.E.S.S. at
a higher energy threshold with data taken at about the same time as MAGIC in
the 2005 period \cite{hess1553}. The source, therefore, can now be considered
as a firmly established VHE $\gamma$-ray emitter.

An interesting aspect of the source is that attempts to determine its redshift
from the optical observations failed so far. However, the VHE $\gamma$-ray
spectrum of PG 1553+113  can be used to derive an upper limit on the redshift
of the source.  Given the observed spectrum of PG 1553+113 and the minimum
possible density of the evolving EBL (lower limit in \cite{kneiske}) we vary
the distance of the source ($z$) until the fit on the reconstructed intrinsic
spectrum yields a photon index, which is beyond the allowed limit ($\alpha <
1.5$, see also Section~\ref{sec:6}). Taking into account the statistical error
on the fit and the systematic error on the slope (0.2) we derived an upper limit
on the redshift of $z < 0.78$. This value is compatible to the one reported in
\cite{hess1553} ($z < 0.74$).  For details of the analysis and results see
\cite{magic1553}.

An alternative method can be used to derive an upper limit on the redshift of
PG 1553+113. The method is based on the indication that assuming rather
large redshifts ($z > 0.3$) the intrinsic spectrum of PG 1553+113 
seems to have a second component above 200 GeV.
To prove this hypothesis (presence of a second component in the PG 1553+113
spectrum) we performed two different statistical tests on the reconstructed
intrinsic spectrum.  In the first test, we fitted the intrinsic spectrum with a
pure power law (fit1) and with a broken power law (fit2). Then we used a {\it
likelihood ratio test} \cite{stat1} to determine the significance that fit2 was a
better hypothesis than fit1 and not just by a coincidence.  As the second test,
a {\it run test} \cite{stat1} was performed on the fit by a pure power law to the
intrinsic spectrum. The {\it run test} was modified with the 
assumption of an equal number of data points below and above the fit.
In order to increase the statistical power, we used the combined spectrum of
MAGIC and H.E.S.S.  data on PG 1553+113 since the two spectra are in good
agreement in the overlapping energy range.  The reconstructed intrinsic
spectrum was considered to have a significant second component at high energies in
case both tests gave more than 2\,$\sigma$ confidence. An example is shown in
Fig.~\ref{fig:1553spec}, assuming a redshift of $z=0.42$. This is the smallest
redshift where both tests gave more than 2\,$\sigma$ confidence for the presence
of the second component. A detailed publication of this analysis is in
preparation.

It cannot be excluded {\it {a priori}} that there is no second component in the
spectrum of PG 1553+113 above 200 GeV. However, in none of the measured VHE
$\gamma$-ray spectra of extragalactic sources such component was found.
Thus, we conclude that either this is the first time that an evidence of a second
component in a VHE $\gamma$-ray spectrum was found or that the redshift of PG
1553+113 is smaller than 0.42.

%%%%%%%%%%%%%%%%%%%%%%%%%%%%%%%%%%%%%%%%%%%%%%%
%   Conclusion
%%%%%%%%%%%%%%%%%%%%%%%%%%%%%%%%%%%%%%%%%%%%%%%

\section{Conclusion and Outlook}
\label{sec:8}

We gave an overview about outstanding findings based on MAGIC observations of
extragalactic objects.  The number of extragalactic VHE $\gamma$-ray sources
detected by MAGIC is currently seven.  Two of them, 1ES 1218+304 and Mkn~180
have been discovered by MAGIC. PG 1553+113 has been co-discovered
with H.E.S.S.  
In general, the reconstructed de-absorbed spectra seem to be the harder the
further away the sources are, which might be related to an observational bias.
Leptonic models (e.g. \cite{mar}) can describe the data, but there are
exceptions like in case of 1ES\,1959+650.  For detailed modeling of the sources,
extensive multiwavelength campaigns including radio through optical telescopes and
X-ray satellites are inevitable.  In case of PG 1553+113, we conclude that
either the redshift of the source is z$<$0.42 or for the first time a source
with a clear evidence of a second emission component in the VHE $\gamma$-ray
energy range has been observed.  Re-observation of the presented sources as
well as the analysis of further extragalactic objects is ongoing.

\begin{acknowledgements}
We would like to thank the IAC for the excellent working conditions at the ORM
in La Palma. 
The support of the German BMBF and MPG, the Italian INFN, the
Spanish CICYT, ETH research grant TH~34/04~3, and the Polish MNiI grant
1P03D01028 is gratefully acknowledged. 
\end{acknowledgements}


\end{document}